\documentclass[praplied,aps,showpacs,superscriptaddress,twocolumn,notitlepage]{revtex4-2}

\usepackage{amsmath,amsfonts,amssymb,amsthm,graphics,graphicx,epsfig,bbm}
\usepackage[colorlinks=true,citecolor=blue,linkcolor=blue,urlcolor=blue]{hyperref}
\usepackage[usenames]{color}
\usepackage{graphicx}
\usepackage{subfigure}
\usepackage{amsmath}
\usepackage{epsfig}
\usepackage{dcolumn}
\usepackage{bm}
\usepackage{color}
\usepackage{times}
\usepackage{epstopdf}
\usepackage{amssymb}
\usepackage{amstext}
\usepackage{latexsym}
\usepackage{hyperref}
\usepackage{amsfonts}
\usepackage{psfrag}
\usepackage{soul,xcolor}
\usepackage[normalem]{ulem}
\usepackage{dsfont}
\usepackage{adjustbox}

\newcommand{\ket}[1]{\vert #1 \rangle}

\newcommand{\ketbra}[2]{\vert #1 \rangle \langle #2 \vert}

\begin{document}

\title{Shortcuts to Adiabaticity for Fast Qubit Readout  in Circuit Quantum Electrodynamics}

\author{F. A. C\'ardenas-L\'opez}
\affiliation{International Center of Quantum Artificial Intelligence for Science and Technology (QuArtist)
and \\  Physics Department, Shanghai University, 200444 Shanghai, China} 

\author{Xi Chen}
\email{chenxi1979cn@gmail.com}
	\affiliation{Department of Physical Chemistry, University of the Basque Country UPV/EHU, Bilbao, Spain}
	\affiliation{EHU Quantum Center, University of the Basque Country UPV/EHU, 48940 Leioa, Spain}

\begin{abstract}

We propose how to engineer the longitudinal coupling to accelerate the measurement of a qubit longitudinally coupled to a cavity, motivated by the concept of shortcuts to adiabaticity. Different modulations are inversely designed from inverse engineering, counter-diabatic driving and genetic algorithm, for achieving optimally large values of the signal-to-noise ratio (SNR) at nanosecond scale. By comparison,
we demonstrate that our protocols outperform the usual periodic modulations on the pointer state separation and SNR. Finally, we show a possible implementation considering state-of-the-art circuit quantum electrodynamics architecture, estimating the minimal time allowed for the measurement process. 

\end{abstract}
\maketitle

\section{Introduction}
Retrieving information from a quantum system is at the heart of quantum information processing applications, in which an accurate and reliable quantum measurement is requisite. The most common measurement strategy is dispersive readout consisting in coupling an auxiliary system whose observables depend on the system state. Specifically, in superconducting quantum circuit (SC) or circuit quantum electrodynamics (cQED)~\cite{Devoret2005book,You2005,Clarke2008,Wendin2005,Devoret2013,Kockum2019,Krantz2019,Kjaergaard2020,Martinis2020,Phys.Rev.A.69.062320,Nature.431.162,Nature.431.159,Nature.451.664,arXiv.2005.12667,Blais2020}, the qubit measurement is carried out through an auxiliary oscillator whose frequency relies on the qubit state~\cite{PhysRevLett.112.190504,PhysRevLett.95.060501,nphys1400,PhysRevA.82.012329,Motzoi2018}. However, dispersive readout has the drawback that its performance is limited by the detuning between the qubit and the oscillator which bounds the number of thermal photons and induces losses via Purcell effect as well~\cite{PhysRevLett.101.080502,PhysRevA.79.013819,PhysRevLett.109.153601}. A way to circumvent this limitation relies on engineer the longitudinal interaction between the quantum systems~\cite{PhysRevA.80.032109,PhysRevB.91.094517,PhysRevLett.115.203601,PhysRevA.95.052333} yielding the non-demolition quantum measurement that is faster and more robust than the previous dispersive approach. Moreover, the parametric modulation of the external magnetic flux on a cavity-qubit system leads to rapid and unconditional reset mechanism~\cite{fastreset}.

\par

In the past decade, shortcuts to adiabaticity (STA) \cite{PhysRevLett.104.063002}  have experienced a huge development, with the various applications of quantum computing and more generally quantum technologies \cite{RevModPhys.91.045001}.  The methods of STA provide efficient control of quantum systems, by accelerating the slow adiabatic processes, and overcoming obstacles from systematic errors or environmental noise, see review \cite{RevModPhys.91.045001}.  Recently, STA have been generalized to complex open systems \cite{PhysRevLett.122.250402,Quantum}, which offer the opportunity for designing a counter-diabatic pulse, to accelerate a dissipative process in cQED \cite{AnNC}. 

In this article, we propose STA method for elaborating the modulation of longitudinal coupling between a two-level system with a cavity mode to accelerate the qubit  measurement. By using inverse engineering, we obtain large values for the cavity pointer state separation and the signal-to-noise ratio (SNR) on a short timescale. 
Besides, the SNR is exponentially enhanced when the cavity is prepared in a squeezed state. For completeness,  we further discuss the measurement process speed-up by counter-diabatic driving, and their possible implementations. Moreover, we use genetic algorithm, a search-based optimization technique, to find optimal or near-optimal  modulation.  Finally, a feasible experimental model regarding state-of-the-art cQED architecture is considered and the minimal time for the measurement process is estimated with the bound of coupling strength.
\par 

\section{Longitudinal cavity-qubit interaction}
We consider an LC oscillator of frequency $\omega_{r}$ longitudinally coupled to a two-level system of frequency $\omega_{q}$ with time-dependent coupling strength $g_{z}(t)$ described through the Hamiltonian~\cite{PhysRevLett.115.203601} ($\hbar =  1$)
\begin{eqnarray}
\label{H_readout}
\mathcal{H} &=&\frac{\omega_{q}}{2}\sigma^{z} + \omega_{r}\hat{a}^{\dag}\hat{a} +  g_{z}(t)\sigma^{z}(\hat{a}^{\dag}+\hat{a}).
\end{eqnarray}
Here $\sigma^{z}$ is the $z-$component Pauli matrix describing the two-level system, and $a^{\dag}$ ($a$) is the creation (annihilation) operator of the LC oscillator. This Hamiltonian corresponds to a state-dependent displaced oscillator. When the cavity acts as a pointer state, we can perform high-fidelity quantum measurement on the qubit state since the cavity state displaces upwards or downwards on its phase state according to the qubit state. Furthermore, as the longitudinal interaction commutes with the free terms of the Hamiltonian such process is a quantum nondemolition measurement (QND)~\cite{PhysRevLett.115.203601}.

We aim to engineer $g_{z}(t)$ inversely to accelerate the QND measurement process. For doing so, we propose a solution for the Sch\"odinger equation of $\mathcal{H}$ as $\ket{\Psi(x,t)} =e^{-iE_{\rm{LC}}t/\hbar}\mathcal{V}(t)\ket{\varphi(x,t)}\ket{\xi}$, where $E_{\rm{LC}}=\omega_{r}(n+1/2)$ and $\ket{\varphi(x,t)}$ correspond to the eigenenergies and eigenfunctions of the LC oscillator $\mathcal{H}_{\rm{LC}}=\omega_{r}\hat{a}^{\dag}\hat{a}$. Besides, $\ket{\xi}$ describes the qubit state. As a consequence,  $\mathcal{V}(t)=e^{i\theta(t)}e^{-i\dot{g}_{c}(t)\sigma^{z}(a^{\dag}+a)/\omega_{r}^{2}}e^{-g_{c}(t)\sigma^{z}(a^{\dag}-a)/\omega_{r}}$ is a unitary transformation that eliminates the longitudinal coupling~\cite{PhysRevLett.112.150402}, where $\theta(t) = -\int_{0}^{t}\mathcal{L}_{g}(t') dt'$ corresponds to a phase relating the coupling strength $g_{z}(t)$ with the auxiliary variable $g_{c}(t)$ through the Lagrangian
\begin{eqnarray}
\label{phases_lagrangian}
\mathcal{L}_{g}(t) &=&\frac{\dot{g}^{2}_{c}(t)}{\omega_{r}^{3}} - \frac{g^{2}_{c}(t)}{\omega_{r}} + \frac{2g_{c}(t)g_{z}(t)}{\omega_{r}}.
\end{eqnarray}
To guarantee that $\ket{\Psi(x,t)}$ corresponds to the exact solution of the time-dependent Sch\"odinger equation, the classical variables must obey the equation of motion, (see Appendix \ref{7} for the detailed calculation)
\begin{eqnarray}
\label{ELE}
\ddot{g}_{c}(t) + \omega_{r}^{2}[g_{c}(t)-g_{z}(t)]=0,
\end{eqnarray}
which is nothing but the Euler-Lagrange equation from Eq.~(\ref{phases_lagrangian}). In what follow, we use Eq. (\ref{ELE}) to engineer inversely the modulation of suitable coupling strength $g_z(t)$, in order to accelerate measurement process.
\par 
\begin{figure}[t!]
\centering
\includegraphics[width=1\columnwidth,height=7cm]{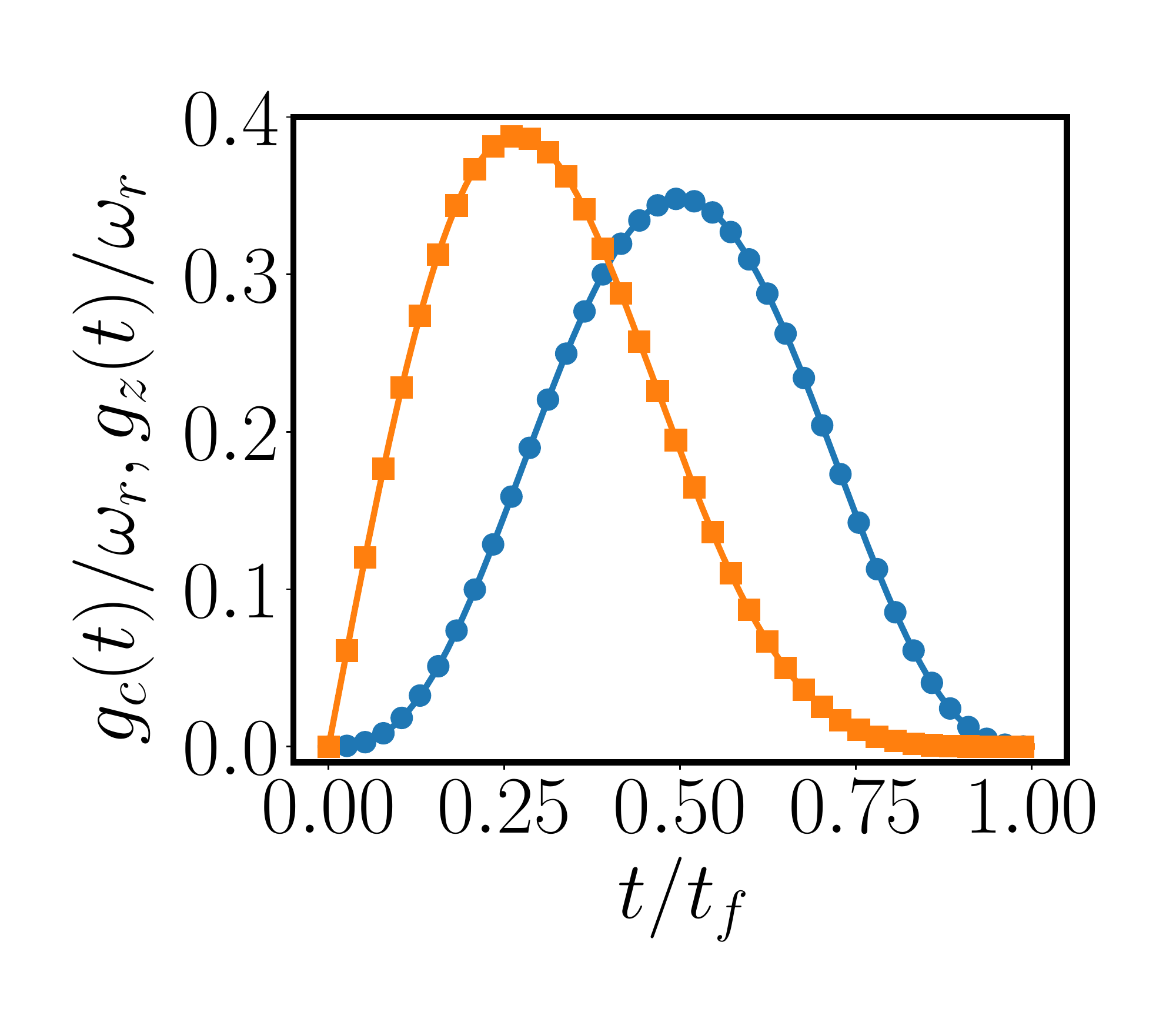}
\caption{Coupling strength modulation $g_{c}(t)$ and $g_{z}(t)$ calculated with the reserve engineering approach. Blue and orange solid line stand for $g_{c}(t)$ for the polynomial and trigonometric ansatzes, respectively, whereas blue dots and orange squares corresponds to $g_{z}(t)$. We have performed the simulation with the parameters $\kappa/2\pi=1~{\rm{MHz}}$, $g_{z0}/2\pi=21~{\rm{MHz}}$, $\omega_{r}/2\pi=6.6~{\rm{GHz}}$ for a designed time $t_{f}=\pi/(100\kappa)$ \cite{PhysRevLett.115.203601}.}
\label{fig:fig1}
\end{figure}

Sharing the concept of STA \cite{PhysRevLett.104.063002}, we require that $g_{c}(t)$ fulfill the following boundary conditions~\cite{PhysRevA.97.013631}:
\begin{subequations}
\begin{eqnarray}
\label{BC0}
g_{c}(0)=0; ~g_{c}(t_{f})=0, ~~~~~~ \\
\dot{g}_{c}(0)=\ddot{g}_{c}(0)=\dot{g}_{c}(t_{f})=\ddot{g}_{c}(t_{f})=0.
\label{BC1}
\end{eqnarray}
\end{subequations}
The flexibility left in  the inverse-engineering approach permits us to add a constrain over the final cavity displacement i.e., 
\begin{eqnarray}
\label{BC2}
\int_{0}^{t_{f}}g_{c}(s) ds= g_{z0}\pi/(2\kappa).
\end{eqnarray}
There exist a vast number of function fulfilling the above criteria, we suggest a polynomial ans\"atze of the form $g_{c}(t)=\sum_{\ell=0}^{6}b_{\ell}t^{\ell}$ that leads to
\begin{eqnarray}
\label{sol1}
g_{c}(t)&=&-\frac{70\pi g_{z0} t^3 (t-t_{f})^3}{\kappa t_{f}^7}.
\end{eqnarray}
However, this is not the unique solution for Eq.~(\ref{ELE}) with the initial and final boundary conditions, see Eqs. (\ref{BC0}), (\ref{BC1}) and (\ref{BC2}). For the generality, we assume  an alternative trigonometric ans\"atze of the form  $g_{c}(t) = \sum^{6}_{m}\mathcal{A}_{m}(t)\sin(m\pi t/t_{f})$, resulting in 
\begin{eqnarray}
\label{sol2}
g_{c}(t) = \frac{3g_{z0}\pi^2}{2\kappa t_{f}}\sin \left(\frac{\pi t}{2t_{f}}\right) \cos ^5\left(\frac{\pi t}{2t_{f}}\right).
\end{eqnarray}
Consequently, we obtain $g_{z}(t)$ by substituting $g_{c}(t)$ into Eq.~(\ref{ELE}), as illustrated in Fig.~\ref{fig:fig1}, where the experimental parameters in an realistic cQED~\cite{PhysRevLett.115.203601} yield $g_z(t) \simeq g_c (t)$ for large frequency $\omega_{r}$ of LC oscillator, 
and the measuring time $t_f$ can be shortened if the larger coupling strength is allowed. 
\begin{figure}[!t]
\centering
\includegraphics[width=1\columnwidth]{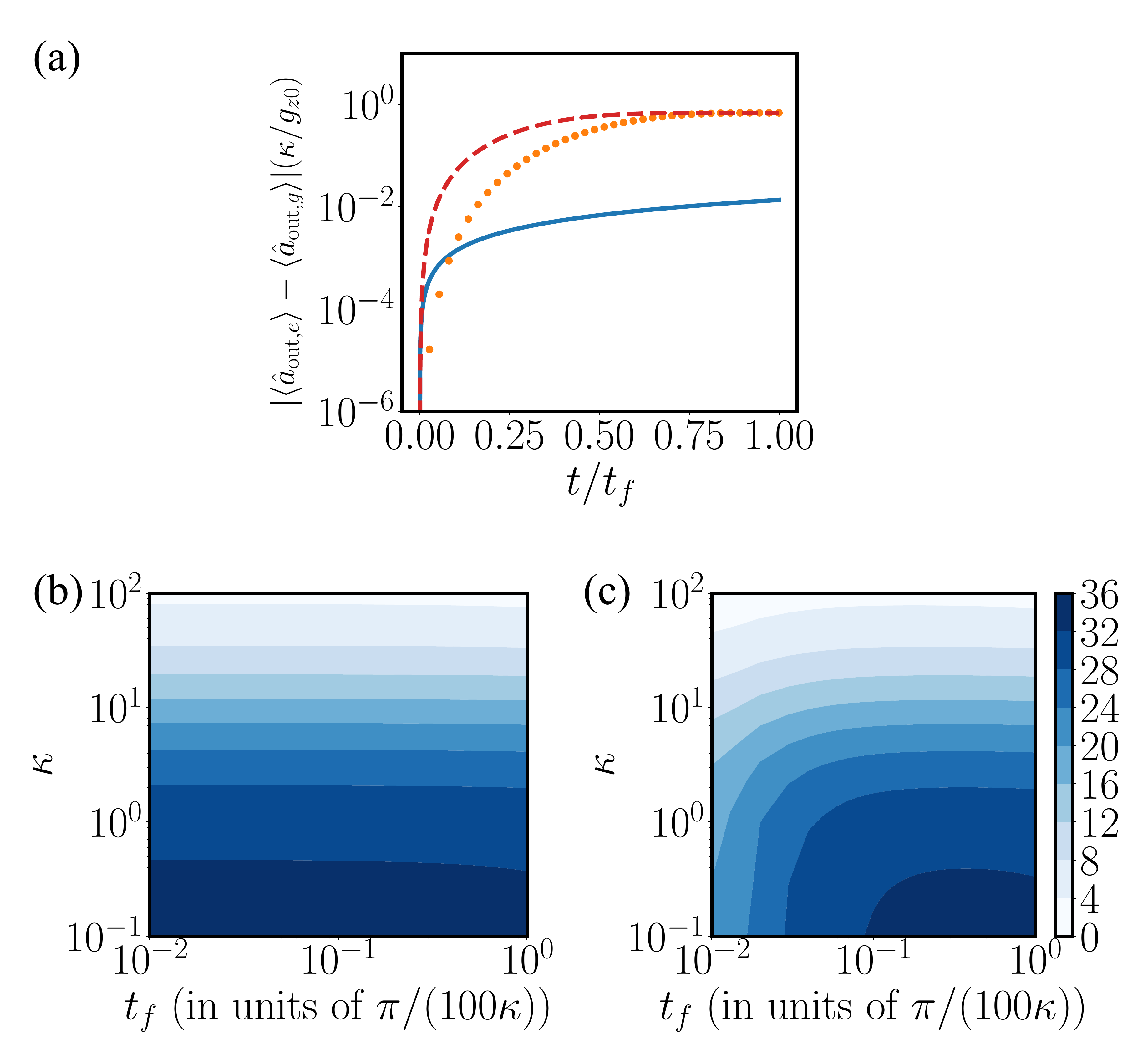}
\caption{(a) Relative pointer state separation for the output cavity field $\hat{a}_{{\rm{out}}}$ as function of $t/t_{f}$ for the inverse engineering with polynomial (orange dotted) and trigonometric (red dashed) ansatzes for longitudinal coupling strength. We compare with the case with the conventional sinusoidal modulation (blue solid).  Function $F(\kappa,t_{f})$ as a function of the decay rate and the final time $t_{f}$ for both polynomial  (b) and trigonometric (c) modulations. All parameters are the same as those in Fig.~\ref{fig:fig2}.}
\label{fig:fig2}
\end{figure}
\section{Qubit readout} 
Now, we consider the evolution of the pointer state with the modulation of  $g_{c}(t)$ given in Eqs.~(\ref{sol1}) and (\ref{sol2}). The dynamical equation of  the cavity field regarding losses~\cite{quantum_noise}  reads $\dot{\hat{a}} = ig_{c}(t)\sigma^{z} - \kappa\hat{a}/2 - \sqrt{\kappa}\hat{a}_{\rm{in}}$, where $\kappa$ is the decay rate of the LC oscillator, and $\hat{a}_{\rm{in}}$ is the input cavity operator taking into account the effect of an additional subsystem (measurement apparatus). By  assuming both $\hat{a}_{\rm{in}}$ and $\hat{a}$ are in their vacuum state, the solution of this equation yields 
\begin{eqnarray}
\label{L_eq3}
\langle\hat{a}(t)\rangle = -i\langle\sigma^{z}\rangle e^{-\kappa t/2}\int_{0}^{t}g_{c}(s)e^{\kappa s/2} ds.
\end{eqnarray}
Depending on the value of $\langle\sigma^{z}\rangle$, the field will be displaced upward or downward on the phase space. To quantify this separation, we  further define the relative separation $d=|\langle\hat{a}_{{\rm{out}},e}(t)\rangle-\langle\hat{a}_{{\rm{out}},g}(t)\rangle|$, where $\langle\hat{a}_{{\rm{out}},\ell}(t)\rangle=\sqrt{\kappa}\langle\hat{a}(t)\rangle$ is the output field averaged with respect to the qubit states $\ket{\ell}=\{\ket{e},\ket{g}\}$. 

Figure~\ref{fig:fig2} shows the dependence of relative separation $d$ on $t/t_{f}$.  Fig.~\ref{fig:fig2} (a) demonstrates that the modulations designed from STA produce the pointer state separation that is ten times larger than the normal sinusoidal modulation in Ref.~\cite{PhysRevLett.115.203601} at timescale $t_{f}=\pi/(100~\kappa)\approx 30~\rm{ns}$, thus yielding a fast qubit readout.  For completeness, we further compare in Fig.~\ref{fig:fig2} (b) the performance of our designed modulations with polynomial and trigonometric ans\"atzes. Though at shorter time trigonometric modulation performs better than polynomial one,  at final time $t_{f}$ both modulations reach the same values, due to the fixed boundary condition (\ref{BC2}). To compare their performance for different $t_{f}$ we plot the quantity $F(\kappa,t_{f})\equiv e^{-\kappa t_{f}/2}\int_{0}^{t_{f}}g_{c}(s)e^{\kappa s/2} ds$ for different $\kappa$ and $t_{f}$. Notice that $F(\kappa,t_{f})$ gives us the maximal cavity displacement at fixed $\kappa$ and $t_{f}$, respectively. From Fig.~\ref{fig:fig2} (b) and Fig.~\ref{fig:fig2} (c), we observe both modulations start to behave similarly, as long as $t$ approaches $t_f$.  Actually, we can also check the fluctuation on the cavity in the final readout.
Based on it, we cannot attribute the enhanced performance to different engineered pulses, which only depends on the final boundary conditions. 


\begin{figure}[!t]
	\centering
	\includegraphics[width=1\columnwidth]{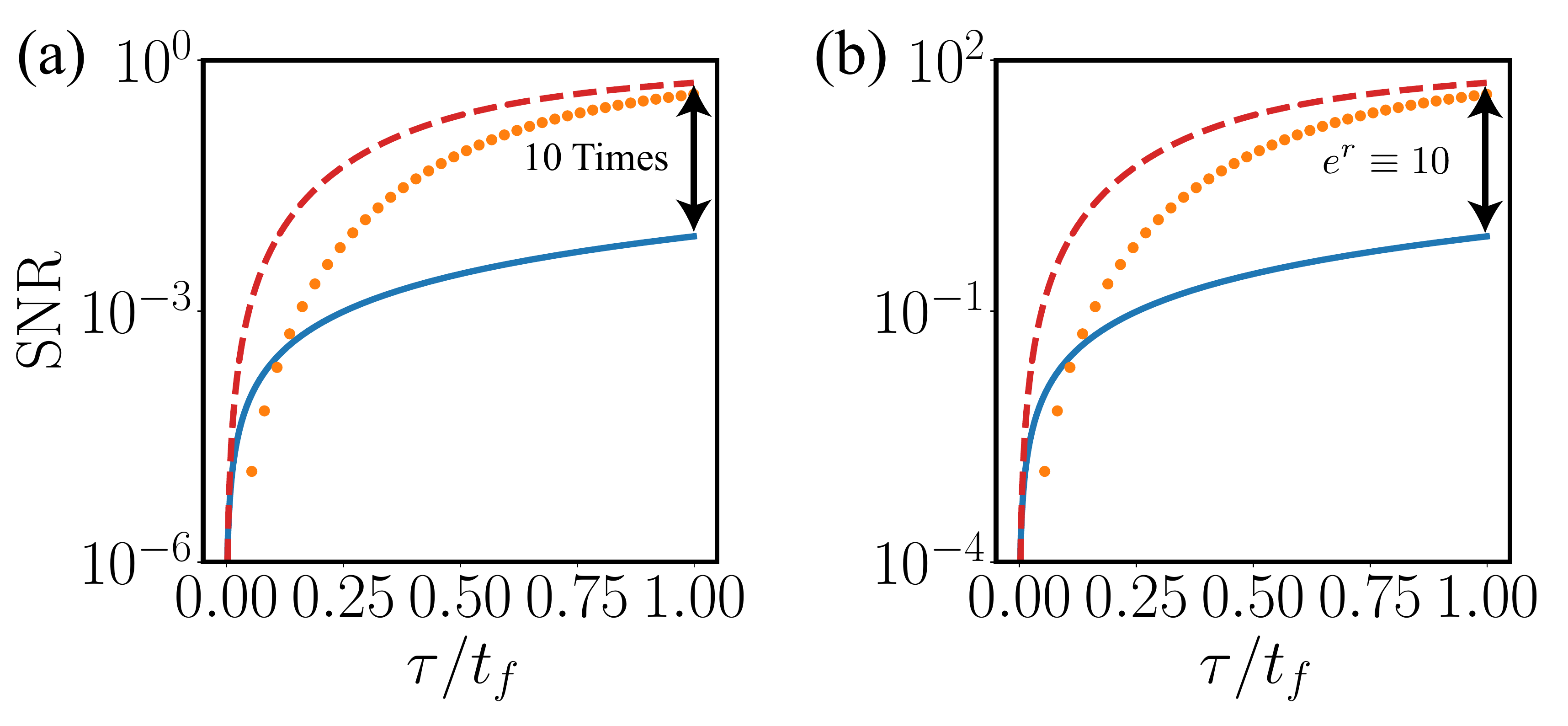}
	\caption{ Signal-to-noise ratio (SNR) as function of the dimensionless measuring time $\tau/t_{f}$ (a) regarding both sinusoidal modulation (blue  solid) and the  inverse-engineered protocol with polynomial (orange dotted) and trigonometric (red dashed) ansatzes. In (b) we also plot the SNR but now considering the cavity with a single-mode squeezed state with squeezing parameter $\theta=\pi/4$ and $r=20~(\rm{dB})\equiv e^{2r}=100$ and homodyne angle $\phi=\pi/4$ for a time $t_{f}=30~\rm{ns}$. The other system parameters for the calculation are the same as those in Fig. \ref{fig:fig2}.}
	\label{fig:fig3}
\end{figure}

Moreover, distant pointer state permits us to perform high-fidelity qubit measurement quantified through the signal-to-noise ratio (SNR) corresponding to the ratio between the homodyne signal with its fluctuations. We define the signal as $|\langle\hat{\mathcal{M}}_{e}\rangle - \langle\hat{\mathcal{M}}_{g}\rangle|$ where $\langle\hat{\mathcal{M}}_{k}\rangle$ is the average of the homodyne operator $\hat{\mathcal{M}}(\tau)=\sqrt{\kappa}\int_{0}^{\tau}ds( a^{\dag}_{{\rm{out}}}(t)\exp(i\phi) + a_{{\rm{out}}}(t)\exp(-i\phi))$ with respect to the qubit state, and we refer to fluctuations by $\sqrt{\langle\hat{\mathcal{M}}_{{\rm{N}}e}^{2}\rangle + \langle\hat{\mathcal{M}}_{{\rm{N}}g}^{2}\rangle}$ with $\hat{\mathcal{M}}_{{\rm{N}}\ell}=\hat{\mathcal{M}}_{\ell}-\langle\hat{\mathcal{M}}_{\ell}\rangle$ as the noise homodyne operator. The SNR is then defined as  
\begin{eqnarray}
\label{SNR}
{\rm{SNR}}(\kappa\tau) = \frac{|\langle\hat{\mathcal{M}}_{e}\rangle - \langle\hat{\mathcal{M}}_{g}\rangle|}{\sqrt{\langle\hat{\mathcal{M}}_{{\rm{N}}e}^{2}\rangle + \langle\hat{\mathcal{M}}_{{\rm{N}}g}^{2}\rangle}}.
\end{eqnarray}
In Fig.~\ref{fig:fig3} (a), we plot the SNR as a function of the dimensionless integration time $\tau/t_{f}$ using the polynomial (red dashed) and trigonometric (orange dotted) modulations, see Eqs.~(\ref{sol1}) and (\ref{sol2}), designed from inverse engineering approach. We see an enhancement in the SNR, achieving the values approximately ten times larger than the sinusoidal modulation proposed at the time scale $t_{f}$. Furthermore, at short measuring time $\kappa \tau \ll1$ we obtain the asymptotic scaling of SNR  as ${\rm{SNR}}(\kappa \tau)\approx(\kappa\tau)^{9/4}$, which approves the enhancement in time on quantum nondemolition measurement. Our result agrees with the statement in Ref.~\cite{PhysRevLett.115.203601} that the readout performance can be improved by quantum control methods.

Moreover, we obtain the further improvement of SNR by considering the LC oscillator prepared in a single-mode squeezed orthogonal to the field displacement. The squeezed state only modifies the noise homodyne operator as $\hat{\mathcal{M}}_{{\rm{N}}\ell}=\kappa\tau(\cosh(2r)+\sinh(2r)\cosh(2(\phi-\theta)))$~\cite{PhysRevLett.115.203601}, where $r$ and $\theta$ are the squeeze parameters and $\phi$ is the homodyne angle. By 
choosing $\phi-\theta=\pi/2 \mod \pi$,  we finally achieve $\hat{\mathcal{M}}_{{\rm{N}}\ell}=\kappa\tau\exp(-2r)$, that leads to the exponential improvement on the SNR illustrated in Fig.~\ref{fig:fig3} (b).

\section{Comparision to Counter-diabatic driving} 

An alternative way to accelerate the qubit readout relies upon the counter-diabatic driving~\cite{RevModPhys.91.045001}. 
Similar to STA for Rabi model \cite{PhysRevLett.126.023602},  the counter-diabatic term for the Hamiltonian in Eq.~(\ref{H_readout})  is calculated
as
\begin{eqnarray}
\mathcal{H}_{\rm{CD}}=-i\frac{\dot{g}_{z}(t)}{\omega_{r}}\sigma^{z}(a^{\dag}-a).
\end{eqnarray} 
This interaction could be implemented in cQED architecture with a tunable capacitive interaction~\cite{Tun_cap_cou}, which has not been easily implemented yet. To circumvent this problem, we utilize the multiple Schr\"{o}dinger/interaction pictures~\cite{PhysRevLett.109.100403} and express the Hamiltonian $\mathcal{H} + \mathcal{H}_{\rm{CD}}$ in a rotating frame, which has  different structure, but  the same underlying physics. By using $\mathcal{U}(t)=\exp(-i\dot{g}_{z}(t)\sigma^{z}(a^{\dag}+a)/\omega_{r}^{2})$, we arrive at the effective Hamiltonian $\bar{\mathcal{H}}=\omega_{q}\sigma^{z}/2 + \omega_{r}\hat{a}^{\dag}\hat{a} +  \tilde{g}_{z}(t)\sigma^{z}(\hat{a}^{\dag}+\hat{a})$, where $\tilde{g}_{z}(t)=g_{z}(t)+\ddot{g}_{z}(t)/\omega_{r}^{2}$ is the effective longitudinal coupling. In this new frame, it only requires a new modulation on the coupling strength $\tilde{g}_{z}(t)$ rather than complicated implementation, for instance, in a current experiment on open cQED \cite{AnNC}.  

Moreover,  such approximate counter-diabatic driving can be implemented by using Floquet engineering (FE)~\cite{Seel2017,Claeys2019,PhysRevResearch.2.013283,PhysRevResearch.3.013227}.  Here, we add a high-frequency driving with a complex time dependency to emulate the counter-diabatic term only using the operators available on the system Hamiltonian. In what follows we will calculate the FE Hamiltonian such that its dynamics corresponds to the same as $\mathcal{H}_{\rm{CD}}$. Essentially, to obtain the state preparation, up to the phase factor, it is not necessary to include the original Hamiltonian $\mathcal{H}$. For doing so, we define the Floquet engineering Hamiltonian as follows:
\begin{eqnarray}
	\mathcal{H}_{\rm{FE}}(t)=\Omega\nu\sin(\nu t)(\sigma^{z}+a^{\dag}a)+\lambda(t)\sigma^{z}(a^{\dag}+a),~~
	\label{HFE}
\end{eqnarray}
where $\nu\ll\omega_{r}$ is an arbitrary frequency, and $\Omega$ is a free parameter. Next, we express $\mathcal{H}_{\rm{FE}}$ in the rotating frame described by the following unitary transformation 
\begin{eqnarray}
	\hat{U}(t)=\exp\bigg[i\Omega\cos\nu t(\sigma^{z} + a^{\dag}a)\bigg].
\end{eqnarray}
In the rotating frame, the effective Hamiltonian $	\tilde{\mathcal{H}}_{\rm{FE}}(t)=\hat{U}^{\dag}(t)\mathcal{H}_{\rm{FE}}\hat{U}(t)-i\hat{U}^{\dag}(t)\dot{\hat{U}}(t)$ reads 
\begin{eqnarray}
	\tilde{\mathcal{H}}_{\rm{FE}}=\lambda(t)\hat{U}^{\dag}(t) \sigma^{z}(a^{\dag}+a)\hat{U}(t).
	\label{HFE2}
\end{eqnarray}
Using the Bakker-Campbell-Hausdorff formula, we express the transformed Hamiltonian as
\begin{eqnarray}
	\tilde{\mathcal{H}}_{\rm{FE}}(t)=\lambda(t)\sigma^{z}(a^{\dag}e^{-i\Omega\cos\nu t}+ae^{i\Omega\cos\nu t}).
	\label{HFE3}
\end{eqnarray}
We proceed by calculating the average of the Hamiltonian over the period $T=2\pi/\nu$ to get the first term of the Magnus expansion
\begin{eqnarray}
	\tilde{\mathcal{H}}^{(0)}_{\rm{FE}}(t)= \frac{1}{T}\int_{0}^{T}\lambda(t)\sigma^{z}(a^{\dag}e^{-i\Omega\cos\nu t}+ae^{i\Omega\cos\nu t})dt.
	\label{HFE4}
\end{eqnarray}
To obtain that $\tilde{\mathcal{H}}^{(0)}_{\rm{FE}}=\mathcal{H}_{\rm{CD}}$ we require that
\begin{eqnarray}
	\frac{1}{T}\int_{0}^{T}\lambda(t)e^{-i\Omega\cos\nu t}dt&=&-i\frac{\dot{g}_{z}(t)}{\omega_{r}},\\
	\label{Cond1}
	\frac{1}{T}\int_{0}^{T}\lambda(t)e^{i\Omega\cos\nu t}dt&=&i\frac{\dot{g}_{z}(t)}{\omega_{r}}.
	\label{Cond2}
\end{eqnarray}
We proceed by assuming that $\lambda(t)=\sum_{n}C_{n}\cos( n \nu t)$, and we define $\alpha=\nu t$. In this case, the integrals above defined read
\begin{eqnarray}
	\sum_{n}\frac{C_{n}}{\pi}\int_{0}^{\pi}\cos(n\alpha)e^{-i\Omega\cos\alpha}d\alpha&=&-i\frac{\dot{g}_{z}(t)}{\omega_{r}}.
	\label{Cond3}
\end{eqnarray}
Notice that this integral looks similar to the definition of the Bessel function
\begin{eqnarray}
	J_{n}(z)=\frac{i^{-n}}{\pi}\int_{0}^{\pi}e^{iz\cos\theta}\cos(n\theta)d\theta.
\end{eqnarray}
With this definition Eq.~(\ref{Cond1}) and Eq.~(\ref{Cond2}) read
\begin{eqnarray}
	\sum_{n}i^{n}C_{n}J_{n}(\Omega)&=&-i\frac{\dot{g}_{z}(t)}{\omega_{r}},\\
	\label{Cond4-a}
	\sum_{n}i^{n}C_{n}J_{n}(-\Omega)&=&i\frac{\dot{g}_{z}(t)}{\omega_{r}}.
	\label{Cond4-b}
\end{eqnarray}
These conditions are meet when $n=2m+1$, $\forall m \in  \mathbb{Z}$. Thus, for $n=1$ we have $\lambda(t)=C_{1}\cos(\nu  t)$ leading to 
\begin{eqnarray}
	C_{1}&=&\frac{\dot{g}_{z}(t)}{\omega_{r}J_{1}(\Omega)}.
	\label{Cond6}
\end{eqnarray}
Finally, the Floquet engineered Hamiltonian takes the following form:
\begin{equation}
\label{HFEF}
\mathcal{H}_{\rm{FE}}(t)=\Omega\nu\sin(\nu t)(\sigma^{z}+a^{\dag}a)+\frac{\dot{g}_{z}(t)}{\omega_{r}J_{1}(\Omega)}\cos(\nu t)\sigma^{z}(a^{\dag}+a), 
\end{equation}
where $\nu \ll \omega_{r}$ is an arbitrary frequency,  $\Omega$ is a free parameter, and $J_{1}$ is the Bessel function of the first kind.  By implementing the only Floquet Hamiltonian (not including the original Hamiltonian $\mathcal{H}$), we show the relative distance between the pointer states and the SNR can be significantly enhanced for different values of $\nu$. 
\begin{figure}[t!]
	\centering
	\includegraphics[width=1\linewidth]{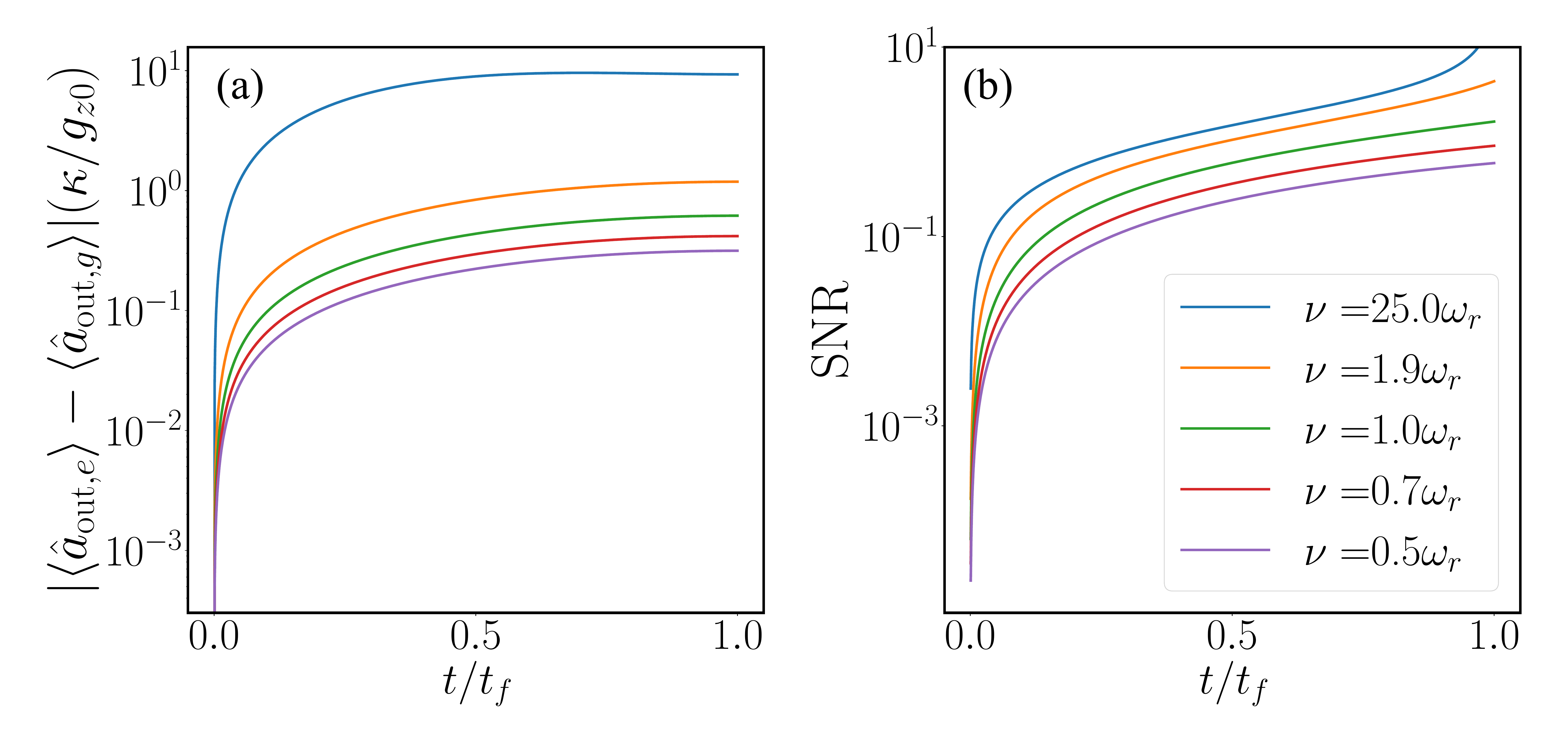}
	\caption{(a) Relative pointer state separation for the output cavity
		field $\hat{a}_{out}$ as function of $t/t_{f}$ for different values of the frequency $\nu$. (b) Signal-to-noise ratio (SNR) as function of the dimensionless measuring time $\tau/t_{f}$. We have performed the simulation with the parameters
		$\kappa/2\pi= 1$~MHz, $g_{z0}/2\pi = 21$~MHz, $\omega_{r}/2\pi = 6.6$~GHz for $t_{f} = \pi/(100\kappa)$, and $\Omega=1$.}
	\label{fig:figA3}
\end{figure}

In Fig.~\ref{fig:figA3}, we show the relative distance between the pointer states and the signal-to-noise ratio (SNR) for the Floquet Hamiltonian in Eq.~(\ref{HFEF}) for different values of $\nu$. As expected by increasing the frequency on the modulation, larger pointer state separation we achieve leading to larger values of the SNR.

\section{Genetic algorithm for optimization}
\begin{figure}[!t]
	\centering
	\includegraphics[width=\columnwidth]{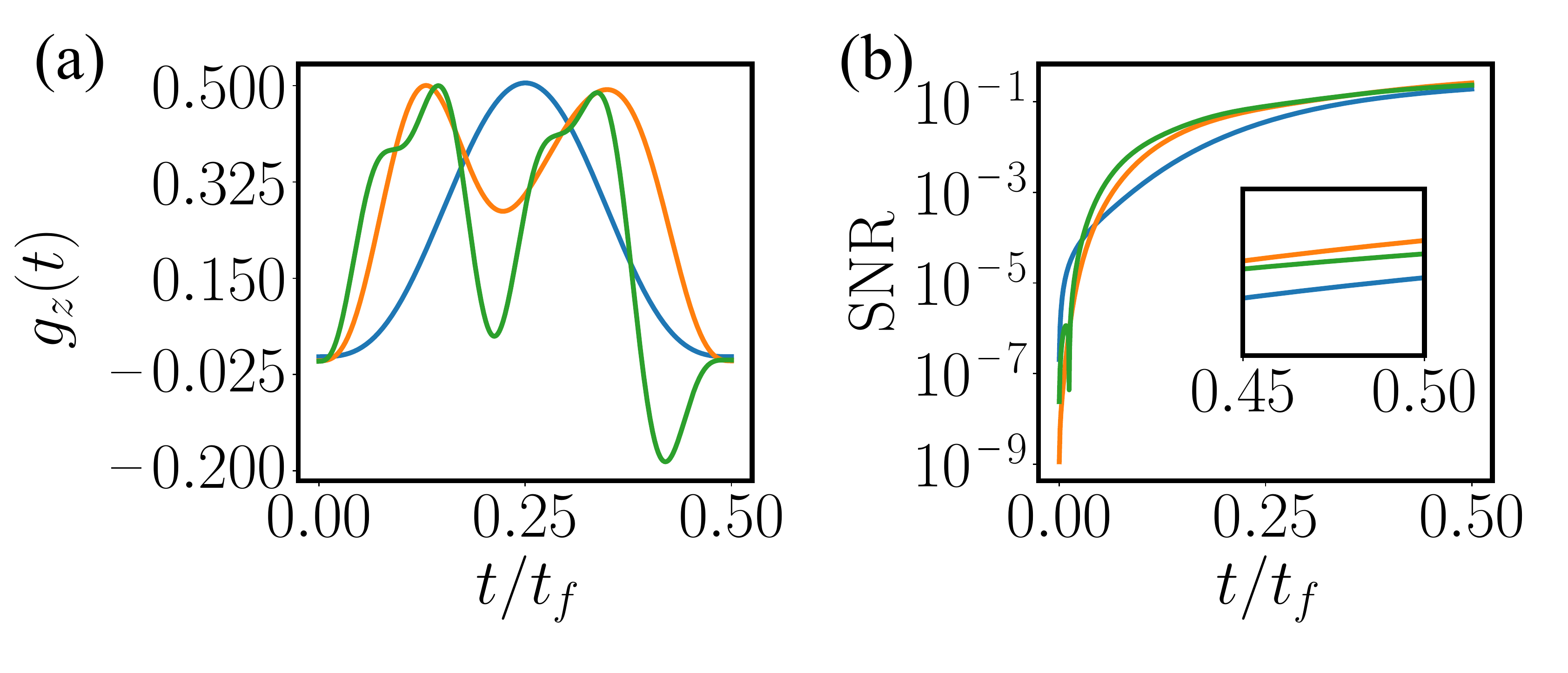}
	\caption{{(a) The modulation of coupling strength $g_{z}(t)$ optimized by the genetic algorithm containing 8 (blue), 12 (orange), and 20 (green) coefficients, respectively. (b) The corresponding SNR obtained with the genetic algorithm. We have performed the simulation with the same parameters as Fig.~\ref{fig:fig2}}, except for the measuring time being $t_f/2$. }
	\label{fig:fig6}
\end{figure}

Now, we turn to the genetic algorithm~\cite{Gen.alg.py}, an optimization subroutine based on the fundamentals of natural selection, in order to complement our inverse-engineering method. We formulate the optimization problem by assuming 
$g_{c}(t)=\sum_{m}c_{m}\cos(m\pi t/t_{f}) + d_{m}\sin(m\pi t/t_{f})$, where the coefficients $\{c_{m}, d_{m}\}$ are to be optimized according the constrains provided by Eqs.~(\ref{BC0}), (\ref{BC1}), and (\ref{BC2}), respectively. Fig.~\ref{fig:fig6} illustrates
the design for $g_{z}$ and corresponding SNR by fixing the number of coefficients. Surprisingly, the increase of the numbers of coefficients does not always lead to better SNR, as depicted in the inline of  Fig.~\ref{fig:fig6} (b), where the performance of 
the modulation with $8$ coefficients surpasses that with 20. In this sense, the genetic algorithms provide a simpler but efficient modulation for
achieving the same SNR at shorter time $t_f/2$.
Of course, other optimization techniques, i.e.,  machine learning~\cite{PhysRevLett.114.200501} and pulse shaping~\cite{PhysRevLett.112.190504,PhysRevApplied.5.011001} can be incorporated as well,  since there exists the freedom left in inverse-engineering method mentioned above.

\section{Physical implementation}

\subsection{Circuit Hamiltonian}
	We shall shed light on the experimental implementation for a two-level system coupled to an oscillator via longitudinal interaction. The circuit consist in an LC resonator of capacitance $C_{r}$ and inductance $L_{r}$ coupled to a transmon qubit~\cite{PhysRevA.76.042319} through a SQUID. The transmon qubit consists of a capacitor $C_{B}$ parallel-connected Josephson junction of capacitance $C_{J}$ and tunable Josephson energy $E_{J}(\phi_{x})$. Moreover, we bias the circuit with an external gate voltage $V_{g}$ connected to the transmon with the gate capacitance $C_{g}$. On the other hand, the SQUID is modeled as a tunable Josephson junction with effective capacitance $C_{JS}$, and Josephson energies $E_{JS}(\varphi_{x})$. We write the Lagrangian of the circuit in terms of the flux nodes of each device $\{\psi_{J},\psi_{{\rm{r}}}\}$ related with the voltage drop across their respective brach $\psi_{\ell}=\int_{-\infty}^{t}V_{\ell}(x,t')dt'$ leading to
	\begin{eqnarray}\nonumber
		\mathcal{L}_{c}&=&\frac{C_{g}}{2}(V_{g}-\dot{\psi}_{J})^2 + \frac{C_{T}}{2}\dot{\psi}_{J}^2 +  \frac{C_{r}}{2}\dot{\psi}_{r}^2\\ \nonumber
		&+&\frac{C_{S}}{2}(\dot{\psi}_{J}-\dot{\psi}_{r})^2-\frac{\psi_{r}^{2}}{2L_{r}}+E_{J}(\phi_{x})\cos\bigg(\frac{\psi_{J}}{\varphi_{0}}\bigg) \\
		&+& E_{JS}(\varphi_{x})\cos\bigg(\frac{\psi_{J}-\psi_{r}}{\varphi_{0}}\bigg),
	\end{eqnarray}
	where $\varphi_{0}=\hbar/(2e)$ is the quantum magnetic flux and $e$ is the electron charge. Besides, $C_{T}=C_{B}+C_{J}$ is the effective transmon capacitance. We calculate the canonical conjugate momenta $P_{\ell}=\partial\mathcal{L}_{c}/\partial[\dot{\psi}_{\ell}]$
	\begin{eqnarray}
		\label{momenta}
		P_{J} &=& (C_{T}+C_{S}+C_{g})\dot{\psi}_{J}-C_{S}\dot{\psi}_{r}-C_{g}V_{g},\\
		P_{r} &=& (C_{r}+C_{S})\dot{\psi}_{r}-C_{S}\dot{\psi}_{J}.
	\end{eqnarray}
\begin{figure}[!t]
	\centering
	\includegraphics[width=\columnwidth]{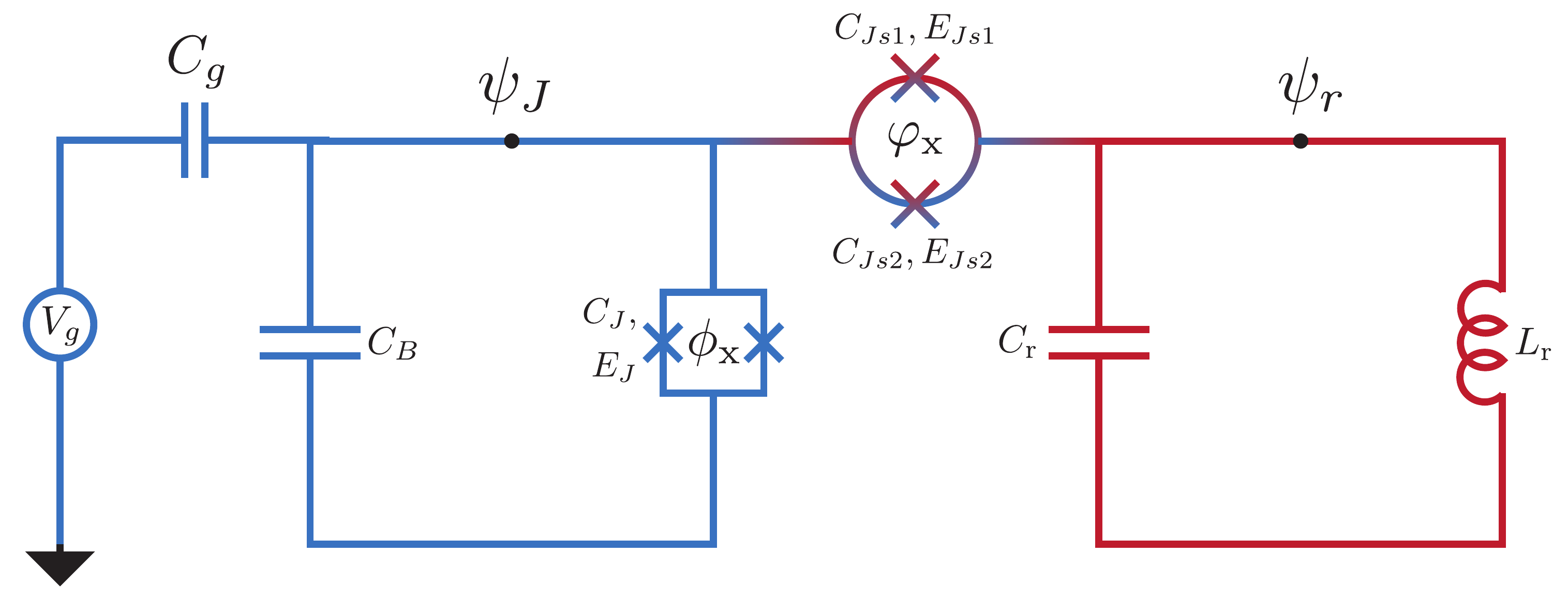}
	\caption{Schematic illustration of the experimental proposal: a transmon qubit formed by a capacitor $C_B$ parallel-connected to a tunable inductor is biased by a gate voltage $V_{g}$ through a gate capacitor $C_{g}$, the two-level system coupled to an LC resonator of capacitance $C_{r}$ and inductance $L_{r}$ via an asymmetric SQUID threaded by an external magnetic flux $\varphi_{x}$. Furthermore, we describe the circuit in terms of their fluxes node $\psi_{J}$ and $\psi_{r}$.}
	\label{fig:fig4}
\end{figure}
Here we have dropped the terms proportional to $V_{g}^{2}$. In matrix form we have $\vec{P}=\hat{\mathcal{C}}~\vec{\Psi} - \vec{Q}_{g}$, where $\vec{P}^{T}=(P_{J},P_{r})$ and $\vec{\Psi}^{T}=(\psi_{J},\psi_{r})$ is the charge and flux vectors, respectively. Furthermore, $\vec{Q}_{g}^{T}=(C_{g}V_{g},0)$ is the gate charge vector, and $\hat{\mathcal{C}}$ is the capacitance matrix 
	\begin{eqnarray}
		\hat{\mathcal{C}}=\begin{pmatrix}
			C_{\Sigma} & -C_{S}  \\
			-C_{S} & C_{\gamma}
		\end{pmatrix},
	\end{eqnarray}
	where $C_{\Sigma}=C_{T}+C_{S}+C_{g}$ and $C_{\gamma}=C_{r}+C_{S}$ are effective transmon and resonator capacitances, respectively. We obtain the circuit Hamiltonian thought the Legendre transformation $\mathcal{H}=\vec{P}^{T}\vec{\Psi}-\mathcal{L}_{c}$ Where $\vec{\Psi}=\hat{\mathcal{C}}^{-1}(\vec{P}+\vec{Q}_{g})$ with $\hat{\mathcal{C}}^{-1}$ beings the inverse of the capacitance matrix. The circuit Hamiltonian reads
	\begin{eqnarray}\nonumber
		\mathcal{H}&=&\frac{C_{\gamma}P_{J}^{2}}{2(C_{\Sigma}C_{\gamma}-C_{S}^2)}+\frac{C_{\Sigma}P_{r}^{2}}{2(C_{\Sigma}C_{\gamma}-C_{S}^2)}+\frac{C_{S}P_{J}P_{r}}{(C_{\Sigma}C_{\gamma}-C_{S}^2)}\\\nonumber
&+&\frac{C_{\gamma}P_{J}Q_{g}}{(C_{\Sigma}C_{\gamma}-C_{S}^2)}+\frac{C_{S}P_{r}Q_{g}}{(C_{\Sigma}C_{\gamma}-C_{S}^2)}+\frac{\psi_{r}^{2}}{2L_{r}}\\
&-&E_{J}(\phi_{x})\cos\bigg(\frac{\psi_{J}}{\varphi_{0}}\bigg)-E_{JS}(\varphi_{x})\cos\bigg(\frac{\psi_{J}-\psi_{r}}{\varphi_{0}}\bigg).
	\end{eqnarray}
	To proceed, we assume that the SQUID works on a parameter regime where the capacitive interaction is smaller than the Josephson energy, regarding only inductive interaction between the subsystems~\cite{PhysRevA.80.032109,PhysRevB.91.094517,PhysRevLett.115.203601,PhysRevA.95.052333}. Thus, for small $C_{S}$ we neglect the capacitive interaction and we rewrite the potential energy obtaining  
	\begin{eqnarray}\nonumber
		\mathcal{H}&=&\frac{C_{\gamma}P_{J}^{2}}{2(C_{\Sigma}C_{\gamma}-C_{S}^2)}+\frac{C_{\Sigma}P_{r}^{2}}{2(C_{\Sigma}C_{\gamma}-C_{S}^2)}+\frac{C_{\gamma}P_{J}Q_{g}}{(C_{\Sigma}C_{\gamma}-C_{S}^2)}\\\nonumber
&+&\frac{\psi_{r}^{2}}{2L_{r}}-E_{J}(\phi_{x})\cos\bigg(\frac{\psi_{J}}{\varphi_{0}}\bigg)\\
&-&E_{JS}(\varphi_{x})\cos\bigg(\frac{\psi_{J}-\psi_{r}}{\varphi_{0}}\bigg).
	\end{eqnarray}
	We now assume that the SQUID works in the linear regime~\cite{PhysRevLett.112.223603} meaning that the mostly of the current flows through the transmon. Hence, the resonator phase is well locate allowing to expand the potential energy up to its leading order in $\psi_{r}/\varphi_{0}$~\cite{Leibjosarr}
	\begin{eqnarray}\nonumber
		&\mathcal{H}&=\frac{C_{\gamma}P_{J}^{2}}{2(C_{\Sigma}C_{\gamma}-C_{S}^2)}+\frac{C_{\Sigma}P_{r}^{2}}{2(C_{\Sigma}C_{\gamma}-C_{S}^2)}+\frac{C_{\gamma}P_{J}Q_{g}}{(C_{\Sigma}C_{\gamma}-C_{S}^2)}\\\nonumber
&-&E_{J}(\phi_{x})\cos\bigg(\frac{\psi_{J}}{\varphi_{0}}\bigg)-E_{JS}(\varphi_{x})\cos\bigg(\frac{\psi_{J}}{\varphi_{0}}\bigg)+\frac{\psi_{r}^{2}}{2L_{r}}\\
		&-&E_{JS}(\varphi_{x})\cos\bigg(\frac{\psi_{J}}{\varphi_{0}}\bigg)\psi_{r}.
	\end{eqnarray}
\begin{figure}[t!]
		\centering
		\includegraphics[width=1\linewidth]{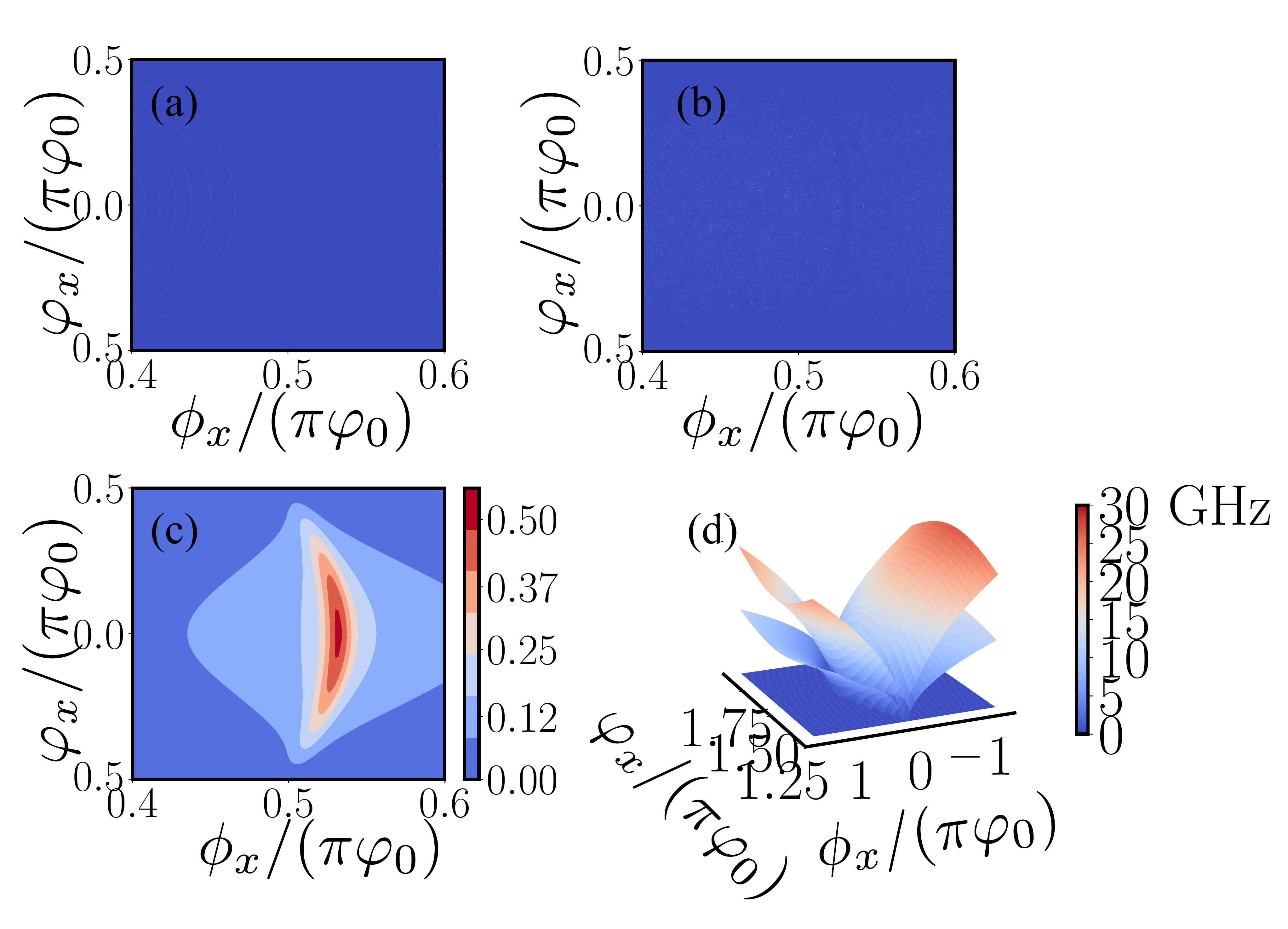}
		\caption{(a)-(c) Pauli matrix coefficient $\alpha_{k}={\rm{Tr}}[\sigma^{k}\cos(\hat{\theta}_{J})]$ for the coupling operator as a function of the external magnetic fluxes $\phi_{x}$ and $\varphi_{x}$. (d) Low-lying energy spectrum of the transmon Hamiltonian $\mathcal{H}_{T}$ as a function of the external magnetic fluxes $\phi_{x}$ and $\varphi_{x}$. We have performed the simulation choosing the parameters $E_{J}/\hbar=2\pi\times20~{\rm{GHz}}$, $E_{C}=E_{J}/67$, $E_{\Sigma}/\hbar=2\pi\times30~{\rm{GHz}}\equiv 1.5~E_{J}$, yielding $\omega_{q}=2\pi\times3.28~{\rm{GHz}}$ and $n_{g}=0.5$.}
		\label{fig:figA2}
	\end{figure}
We proceed to quantize the circuit Hamiltonian by promoting the classical variables to quantum operators. For the transmon qubit the charge of the circuit is proportional to the number of Cooper-pair $P_{J}\rightarrow-2e\hat{n}_{J}$ and its conjugate variable corresponds to the phase drop $\hat{\theta}_{J}=\psi_{J}/\varphi_{0}$ satisfying commutation relation $[\hat{P}_{J},e^{i\hat{\theta}_{J}}]=i$. For the LC resonator, the operators satisfies $[\hat{\psi}_{J},\hat{P}_{J}]=i\hbar$. The quantum circuit Hamiltonian reads
	\begin{eqnarray}
		\label{circuit_hamiltonian-SM-1}\nonumber
		\mathcal{H}&=&E_{C}(\hat{n}_{J}-n_{g})^2  - E_{\tilde{J}}(\phi_{x},\varphi_{x})\cos\big(\hat{\theta}_{J}\big)\\
		&+& \hbar\omega_{r}\hat{a}^{\dag}\hat{a}- \frac{E_{JS}(\varphi_{x})}{\varphi_{0}}\sqrt{\frac{\hbar\omega_{r}L_{r}}{2}}(\hat{a}^{\dag}+\hat{a})\cos\big(\hat{\theta}_{J}\big).~~~~
	\end{eqnarray}
	Here, $E_{C}=2e^2C_{\gamma}/(C_{\Sigma}C_{\gamma}-C_{S}^2)$ and $E_{\tilde{J}}(\phi_{x},\varphi_{x})=E_{J}(\phi_{x})+E_{JS}(\varphi_{x})$ correspond to the charge energy and the effective Josephson energy of the transmon, respectively, $n_{g}=Q_{g}/2e^2$ stands for the dimensionless gate charge. Besides, $\omega_{r}=\sqrt{C_{\Sigma}/((C_{\Sigma}C_{\gamma}-C_{S}^2)L_{r})}$ is the oscillator frequency. It is convenient to divide the circuit Hamiltonian in three parts
	\begin{eqnarray}
		\label{circuit_hamiltonian-SM-2}
		\mathcal{H}_{T}&=&E_{C}(\hat{n}_{J}-n_{g})^2  - E_{\tilde{J}}(\phi_{x},\varphi_{x})\cos\big(\hat{\theta}_{J}\big),\\ 
		\mathcal{H}_{r}&=&\hbar\omega_{r}\hat{a}^{\dag}\hat{a},\\
		\mathcal{H}_{I}&=&- \frac{E_{JS}(\varphi_{x})}{\varphi_{0}}\sqrt{\frac{\hbar\omega_{r}L_{r}}{2}}(\hat{a}^{\dag}+\hat{a})\cos\big(\hat{\theta}_{J}\big),
	\end{eqnarray}
	corresponding to the transmon, resonator and interaction Hamiltonian respectively. 
	
	\subsection{Two-level approximation}
	Next,  we turn to illustrate that in the two-level approximation of the transmon qubit, the Hamiltonian $\mathcal{H}_{I}$ leads to a longitudinal oscillator qubit interaction. We express this Hamiltonian in the charge basis $\ket{n_{J}}$ choosing $\ket{g}\equiv\ket{0}$ and $\ket{e}\equiv\ket{1}$ the Hamiltonian reads (up to terms proportional to $n_{g}^2$
	\begin{eqnarray}
		\label{circuit_hamiltonian-SM-3}
		\mathcal{H}_{T}&=&  E_{C}(1-2n_{g})\ketbra{e}{e} - \frac{E_{\tilde{J}}(\phi_{x},\varphi_{x})}{2}\sigma^{x},\\ 
		\mathcal{H}_{I}&=&- \frac{E_{JS}(\varphi_{x})}{2\varphi_{0}}\sqrt{\frac{\hbar\omega_{r}L_{r}}{2}}(\hat{a}^{\dag}+\hat{a})\sigma^{x},
	\end{eqnarray}
	where $\sigma^{x}=\ketbra{1}{0}+\ketbra{0}{1}$ For a external gate charge $n_{g}=0.5$ we obtain that the first term of $\mathcal{H}_{T}$ vanishes. Then it is possible to write as follows
	\begin{eqnarray}
		\label{circuit_hamiltonian-SM-4}
		\mathcal{H}_{T}&=&\frac{\hbar \omega_{q}(\phi_{x},\varphi_{x})}{2}\sigma^{z},
	\end{eqnarray}
	where $\omega_{q}=\sqrt{E_{\tilde{J}}^{2}(\phi_{x},\varphi_{x})}/\hbar$ is the transition frequency of the qubit. Similar to the coupling operator. To prove that, In Fig.~\ref{fig:figA2}, we have calculated the coefficients $\alpha_{k}={\rm{Tr}}[\sigma^{k}\cos(\hat{\theta}_{J})]$ writing the operator $\cos\big(\hat{\theta}_{J}\big)$ in the diagonal basis of $\mathcal{H}_{T}$ observing that there is no contributions of either $\sigma^{x}$ or $\sigma^{y}$. Therefore, the interaction Hamiltonian can be expressed as follows 
	\begin{eqnarray}
		\mathcal{H}_{I}&=&- \frac{E_{JS}(\varphi_{x})}{2\varphi_{0}}\sqrt{\frac{\hbar\omega_{r}L_{r}}{2}}(\hat{a}^{\dag}+\hat{a})\sigma^{z}.
	\end{eqnarray}
	Finally, the circuit Hamiltonian reads
	\begin{eqnarray}
		\label{H_LC-SM}
		\mathcal{H}&=&\hbar\omega_{r}\hat{a}^{\dag}\hat{a}+\frac{\hbar\omega_{q}}{2}\sigma^{z}+\hbar g_{z}(t)(\hat{a}^{\dag}+\hat{a})\sigma^{z},\\
g_{z}(t)&=&\frac{\omega_{q}}{2\varphi_{0}}\sqrt{\frac{\hbar\omega_{r}L_{r}}{2}}.
	\end{eqnarray}
	In Fig.~\ref{fig:figA2} we have also plot the energy spectrum of the transmon Hamiltonian as function of both external magnetic fluxes $\phi_{x}$ and $\varphi_{x}$. We see that the energy spectrum of the two-level system does not exhibit abrupt changes along $\varphi_{x}$ corresponding to the tunable coupling strength. Consequently, it is possible to switch the coupling strength without modifying the energy spectrum of the qubit.
	
	\subsection{Estimation of the coupling strength and minimal time}
	
	From the last subsection, we have obtained that under suitable conditions the circuit corresponds to a qubit longitudinally coupled to an oscillator with coupling strength given by 
	\begin{eqnarray}
		g_{z}(t)=\frac{\omega_{q}}{2\varphi_{0}}\sqrt{\frac{\hbar\omega_{r}L_{r}}{2}}.
	\end{eqnarray}
	From this expression we observe that the coupling strength depends mainly on four parameters, the qubit frequency, the sum of the Josephson energy of the SQUID, and the resonator capacitance and inductance. Thus considering consistent cQED values it is possible to estimate the maximal value of $g_{z}(t)$. To achieve larger values of the coupling strength that yields to faster measuring time we require large impedance~\cite{PhysRevApplied.5.044004}. In this direction, technological progress has made possible to engineer inductances in the $\mu$H regime using arrays of Josephon junctions or taking into account kinetic inductors~\cite{Pechenezhskiy2020,Andersen2016,Stockklauser2017,Grunhaupt2019}. To estimate the value of the coupling strength we regard $\phi_{x}/\varphi_{0}=\varphi_{x}/\varphi_{0}=\pi/4$, hence the qubit frequency turns into $\omega_{q}=\sqrt{E_{C}^{2}+dE_{\Sigma}^{2}}/\hbar$. For realistic cQED parameters $E_{J}/\hbar=2\pi\times20~{\rm{GHz}}$, $E_{C}=E_{J}/67$~\cite{PhysRevLett.115.203601}, $E_{\Sigma}/\hbar=2\pi\times30~{\rm{GHz}}\equiv 1.5~E_{J}$~\cite{SQUID} and $d=0.02$ we obtain $\omega_{q}=2\pi\times3.28~{\rm{GHz}}$. Moreover, for an LC oscillator (or transmission line resonator) having values  $\omega_{r}L_{r}=200~{{\rm{k}}}\Omega$~\cite{Pechenezhskiy2020} we achieve $g_{z}(t)= 2 \pi \times 2.57~\rm{GHz}$ corresponding to $\max(g_{z})/\omega_{r}\equiv0.5793$.
	
	With this maximal value, it is possible to estimate the minimal time required to measure the qubit $t_{{\rm{min}}}=\pi/(2\omega_{r})$,  on a subnanosecond time scale,  from optimal control theory, see the detailed discussion in Appendix \ref{OCT}.

\section{Conclusion}
In summary, the methods of STA, including inverse engineering and counter-diabaticity, have been worked out for designing the longitudinal qubit-cavity coupling to accelerate the qubit measurement. Remarkably, by engineering the modulations, the pointer state separation is significantly enhanced, accompanied by a large SNR.  We also see an exponential enhancement when the cavity in a single-mode is prepared in squeezed state. In addition,  genetic algorithm are discussed also for the optimization. In the cQED platform, tunable capacitive interaction is required to implement counter-diabatic driving, which makes the inverse-engineering approach more feasible to speed up the measurement process, with the realistic cQED architecture. We estimate  an upper bound for the coupling strength that set the low bound for measuring time. Last but not least, we hope our result can be experimentally verified with circuit design implementing  longitudinal coupling of superconducting qubits \cite{PhysRevLett.115.203601,PhysRevB.93.134501}, and applicable to electronic spin readout as well \cite{PhysRevB.79.041302,PhysRevA.95.012312}.

\section*{ACKNOWLEDGMENTS}
This work is partially supported from NSFC (12075145), STCSM (2019SHZDZX01-ZX04), QMiCS (820505) and OpenSuperQ (820363) of the EU Flagship on Quantum Technologies, EU FET Open Grant Quromorphic (828826) and EPIQUS (899368),  QUANTEK project (KK-2021/00070), and  the Basque Government through
Grant No. IT1470-22. 
X.C. acknowledges the Ram\'on y Cajal program (RYC-2017-22482).

\appendix
\section{Elimination of the longitudinal coupling}\label{7}
Let us consider a two-level system longitudinally coupled to a oscillator described by the Hamiltonian Eq. (\ref{H_readout}),
where $\omega_{q}$ is the transition frequency of the qubit, $\omega_{r}$ is the oscillator frequency. Furthermore, $\sigma^{z}$ is the $z-$component Pauli matrix describing the two-level system, and $a^{\dag}$ ($a$) is the creation (annihilation) operator of the oscillator. We will show that the unitary transformation
\begin{eqnarray}
	\label{unit}\nonumber
	\mathcal{V}(t)=e^{i\theta(t)}e^{-i\frac{\dot{g}_{c}(t)\sigma^{z}(a^{\dag}+a)}{\omega_{r}^{2}}}e^{-\frac{g_{c}(t)}{\omega_{r}}\sigma^{z}(a^{\dag}-a)},
\end{eqnarray}
presented in the manuscript eliminates the longitudinal coupling strength leading to the effective Hamiltonian $\mathcal{H}_{{\rm{eff}}}=\mathcal{V}^{\dag}\mathcal{H}\mathcal{V}-i\dot{\mathcal{V}}\mathcal{V}^{\dag}\equiv \omega_{r}a^{\dag}a+\omega_{q}\sigma^{z}/2$. 
Here, $\theta(t)$ corresponds to a phase defined as 
\begin{eqnarray}
	\label{phase}
	\theta(t) = -\int_{0}^{t}\mathcal{L}_{g}(t') dt,
\end{eqnarray}
where $\mathcal{L}_{g}$ is a Lagrangian that relates the quantities $g_{z}(t)$ and $g_{c}(t)$ through the following relation:
\begin{eqnarray}
	\label{phases_lagrangian-SM}
	\mathcal{L}_{g}(t) &=&\frac{\dot{g}^{2}_{c}(t)}{\omega_{r}^{3}} - \frac{g^{2}_{c}(t)}{\omega_{r}} + \frac{2g_{c}(t)g_{z}(t)}{\omega_{r}}.
\end{eqnarray}
For this derivation it is convenient to divide in three the effective Hamiltonian in three terms regarding the free terms ($\mathcal{H}_{{\rm{eff}},1}$), the transformed longitudinal interaction ($\mathcal{H}_{{\rm{eff}},2}$) and the terms appearing due the transformation $\mathcal{H}_{{\rm{eff}},3}=-i\dot{\mathcal{V}}\mathcal{V}^{\dag}$. For the free terms we obtain
\begin{eqnarray}
	\label{H2}\nonumber
	\mathcal{H}_{{\rm{eff}},1} &=& \mathcal{V}^{\dag}\bigg(\frac{\omega_{q}}{2}\sigma^{z} + \omega_{r}\hat{a}^{\dag}\hat{a}\bigg)\mathcal{V}\\\nonumber
	&=& \frac{\omega_{q}}{2}\sigma^{z} + \omega_{r}\hat{a}^{\dag}\hat{a} + \frac{i\dot{g}_{c}(t)}{\omega_{r}}\sigma^{z}[\hat{a}^{\dag}+\hat{a},\hat{a}^{\dag}\hat{a}]\\ \nonumber
	&+& g_{c}(t)\sigma^{z}[\hat{a}^{\dag}-\hat{a},\hat{a}^{\dag}\hat{a}]-\frac{\dot{g}_{c}^{2}(t)}{2\omega_{r}^{3}}[\hat{a}^{\dag}+\hat{a},[\hat{a}^{\dag}+\hat{a},\hat{a}^{\dag}\hat{a}]]\\\nonumber
	&+&\frac{i\dot{g}_{c}(t)g_{c}(t)}{2\omega_{r}^{2}}[\hat{a}^{\dag}+\hat{a},[\hat{a}^{\dag}-\hat{a},\hat{a}^{\dag}\hat{a}]]\\\nonumber
	&+& \frac{i\dot{g}_{c}(t)g_{c}(t)}{2\omega_{r}^{2}}[\hat{a}^{\dag}-\hat{a},[\hat{a}^{\dag}+\hat{a},\hat{a}^{\dag}\hat{a}]]\\ 
	&+& \frac{g_{c}^{2}(t)}{2\omega_{r}}[\hat{a}^{\dag}-\hat{a},[\hat{a}^{\dag}-\hat{a},\hat{a}^{\dag}\hat{a}]].
\end{eqnarray} 
In this derivation, we have used the Baker-Campbell-Hausdorff formulae keeping terms up to second order in the coupling strength $g_{c}$ and $\dot{g}_{c}$, respectivelly. After solving the commutators we arrive at
\begin{eqnarray}
	\label{H3}\nonumber
	\mathcal{H}_{{\rm{eff}},1} &=& \frac{\omega_{q}}{2}\sigma^{z} + \omega_{r}\hat{a}^{\dag}\hat{a}\\\nonumber
	&-& \frac{i\dot{g}_{c}(t)}{\omega_{r}}\sigma^{z}(\hat{a}^{\dag}-\hat{a}) - g_{c}(t)\sigma^{z}(\hat{a}^{\dag}+\hat{a})\\
	&+&\frac{\dot{g}_{c}^{2}(t)}{\omega_{r}^{3}} + \frac{g_{c}^{2}(t)}{\omega_{r}}.
\end{eqnarray} 
With the same procedure we obtain the transformed longitudinal coupling $\mathcal{H}_{{\rm{eff}},2}$ as follows 
\begin{eqnarray}
	\label{H4}
	\mathcal{H}_{{\rm{eff}},2} =g_{z}(t)\sigma^{z}(\hat{a}^{\dag}+\hat{a})-\frac{2g_{c}(t)g_{z}(t)}{\omega_{r}}.
\end{eqnarray} 
Similar for $\mathcal{H}_{{\rm{eff}},3}$ we obtain 
\begin{eqnarray}
	\label{H5}\nonumber
	\mathcal{H}_{{\rm{eff}},3} &=&-\dot{\theta}(t)+\frac{\ddot{g}_{c}(t)}{\omega_{r}^{2}}\sigma^{z}(\hat{a}^{\dag}+\hat{a})\\
	&+& \frac{i\dot{g}_{c}(t)}{\omega_{r}}\sigma^{z}(\hat{a}^{\dag}-\hat{a}) - \frac{2\dot{g}^{2}_{c}(t)}{\omega_{r}^3}.
\end{eqnarray} 
Finally, we arrive at the effective Hamiltonian
\begin{eqnarray}
	\label{H6}\nonumber
	\mathcal{H}_{{\rm{eff}}} &=& \frac{\omega_{q}}{2}\sigma^{z} + \omega_{r}\hat{a}^{\dag}\hat{a}-\dot{\theta}(t)\\\nonumber
	&+& \bigg[\frac{\ddot{g}_{c}(t)}{\omega_{r}^{2}}-g_{c}(t)+g_{z}(t)\bigg]\sigma^{z}(\hat{a}^{\dag}+\hat{a})\\
	&-&\bigg[\frac{\dot{g}_{c}^{2}(t)}{\omega_{r}^{3}}-\frac{g_{c}^{2}(t)}{\omega_{r}}+\frac{2g_{c}(t)g_{z}(t)}{\omega_{r}}\bigg].
\end{eqnarray} 
From the effective Hamiltonian in Eq.~(\ref{H6}) we see that for achieve $\mathcal{H}_{{\rm{eff}}}=\mathcal{V}^{\dag}\mathcal{H}\mathcal{V}-i\dot{\mathcal{V}}\mathcal{V}^{\dag}\equiv \omega_{r}\hat{a}^{\dag}\hat{a}+\omega_{q}\sigma^{z}/2$, we require satisfy two conditions;
\begin{eqnarray}
	\label{cond1}
	&&\dot{\theta}(t) =\frac{\dot{g}_{c}^{2}(t)}{\omega_{r}^{3}}-\frac{g_{c}^{2}(t)}{\omega_{r}}+\frac{2g_{c}(t)g_{z}(t)}{\omega_{r}},\\
	\label{cond2}
	&&\ddot{g}_{c}(t)+\omega_{r}^{2}[g_{c}(t)-g_{z}(t)]=0.
\end{eqnarray}
Notice that Eq.~(\ref{cond1}) is exactly the same as the Lagrangian in Eq.~(\ref{phases_lagrangian-SM}), and the condition Eq.~(\ref{cond2}) is nothing but the Euler-Lagrange equation for the Lagrangian $\mathcal{L}_{g}$. Thus, we concluded that the unitary transformation $\mathcal{V}(t)$ permit to us to express the system Hamiltonian $\mathcal{H}$, see Eq. (\ref{H_readout}), in a frame without longitudinal coupling strength.

	\section{Time-optimal control }
	\label{OCT}
	  Given the freedom left in reverse engineering based on Eq. (\ref{ELE}), we combine it with optimal control theory (OCT) to find the minimal time for measurement according to the maximal coupling strength obtained above.  In order to account for boundary conditions, we first enlarge the control system as $g_d=\dot{g}_c$. The state of the system $X=(g_c, g_d)^\intercal$ satisfies the following differential system:
	\begin{equation}
		\dot{X}=AX+uB,
	\end{equation}
	with the control $u=g_z$ and the matrices $A$ and $B$ defined as follows:
	\begin{eqnarray}
		A=\begin{pmatrix}
			0 & 1 \\
			-\omega_r^2 & 0 
		\end{pmatrix} ~ \mbox{and}
		~B=\begin{pmatrix}
			0 \\ \omega_r^2 
		\end{pmatrix}.
	\end{eqnarray}
	We reformulate Eq. (\ref{ELE}) into time-optimal control problem, by defining dynamical equations 
	\begin{eqnarray}
		\dot{g}_c &=& g_d,
		\\
		\dot{g}_d &=& -\omega^2_{r} (g_c-u).
	\end{eqnarray}
	To minimize the time $J= \int_{0}^{t_f} dt$, we apply the Pontryagin maximum principle to find the control $\{u, X(t)\}$ under the constraint $ 0  \leq u  \leq u_m$ ($u_m= \max(g_z)$), consistent with the boundary conditions. 
	The optimal control Hamiltonian is $H_c= p_0 + p_g g_d + p_d \omega^2_{r} (u-g_c) $, with $p_0$ being the constant and $(p_g,p_d)$ being the multipliers. The Pontryagin maximum principle tells \cite{Pontryagin} us that the adjoint state 
	is the solution of the following differential equations, 
	\begin{eqnarray}
		\dot{p}_g &=& \omega^{r} p_d,
		\\
		\dot{p}_d &=& -p_g,
	\end{eqnarray}
	from which the adjoint equations can be obtained as 
	\begin{eqnarray}
		p_d (t) = A_d \cos(\omega_r t) +B_d \sin(\omega_r t),
	\end{eqnarray}
	where $A_d$ and $B_d$ can be fixed by initial values $p_d (0)$. 
	We can also introduce the switching function $\Phi=p_d$, such that 
	\begin{align}
		u= \left\{ 
		\begin{array}{rcl}
			\max(g_z),        &    & {\Phi >0} \\
			0,    &    & {\Phi<0} 
		\end{array} \right.,
	\end{align}
	In the singular case we have $p_d=p_g=0$ on a non-zero time interval, and this extremal cannot be reached since ${p_d, p_g}$ are continuous. When $u=0$,
	we have $\ddot{g}_c + \omega^2_r g_c=0$, yielding
	\begin{eqnarray}
		g_c (t) &=& A_0 \cos(\omega_r t) +B_0 \sin(\omega_r t),
		\\
		g_d (t) &=& -A_0 \omega_r \sin(\omega_r t) +B_0 \omega_r \cos(\omega_r t),
	\end{eqnarray}
	where $A_0$ and $B_0$ are constants determined later by boundary conditions. 
	From this, we deduce that $g_d$ and $g_c$ cannot be simultaneously equal to zero at initial or final times, so $u=0$ does not correspond to the first or last bang. 
	However, we can still estimate the minimal time when $u=u_m$ on $[0, t_f]$. Therefore, from the initial boundary condition, we obtain:
	\begin{eqnarray}
		g_c (t) &=& u_m (1-\cos(\omega_r t)),
		\\
		g_d (t) &=&  \omega_r u_m \sin(\omega_r t).
	\end{eqnarray}
	We can notice that $\omega_r t_f = 2 k \pi$ ($k=1,2,3...$), which gives the minimal measurement time $t_{\min} = \pi/(2\omega_{r})$ ($k=1$), which is on the subnanosecond time scale with the system parameters used here.
	Of course, one may consider other optimal problems with various constraints as well.

\end{document}